\documentclass[submission,copyright,creativecommons]{eptcs}

\usepackage{cite}
\usepackage{amsmath,amssymb,amsfonts}
\usepackage{algorithmic}
\usepackage{graphicx}
\usepackage {svg}
\usepackage{textcomp}
\usepackage{xcolor}
\usepackage{tcolorbox}

\usepackage{braket} 

\newcommand{\tabincell}[2]{\begin{tabular}{@{}#1@{}}#2\end{tabular}}

\usepackage{booktabs} 
\usepackage{multirow}
\usepackage{float}
\newfloat{Figure}{htbp}{lof}[section]
\usepackage{subfigure}
\usepackage{diagbox}
\usepackage{enumitem} 

\usepackage{url}
\usepackage{stmaryrd}
\usepackage{cleveref}
\usepackage{amsfonts}
\usepackage{makecell}

\usepackage{times}
\usepackage{graphicx}
\usepackage{epsf}
\usepackage{verbatim}
\usepackage{url}
\usepackage{color}
\usepackage{alltt}

\newcommand{\Comment}[1]{}

\newenvironment{CodeOut}{\begin{scriptsize}}{\end{scriptsize}}

\newcommand{\SmallSpace}{\vspace*{-1.4ex}}

\title{On Code Safety for Quantum Intermediate Representations}
\title{Formalization of Quantum Intermediate Representations\\ for Code Safety}

\author{Junjie Luo
\institute{Kyushu University}
\email{luo.junjie.609@s.kyushu-u.ac.jp}
\and
Jianjun Zhao
\institute{Kyushu University}
\email{zhao@ait.kyushu-u.ac.jp}
}

\def\titlerunning{}
\def\authorrunning{}
\begin{document}
\maketitle

\begin{abstract}
Quantum Intermediate Representation (QIR) is a Microsoft-developed, LLVM-based intermediate representation for quantum program compilers. QIR aims to provide a general solution for quantum program compilers independent of front-end languages and back-end hardware, thus avoiding duplicate development of intermediate representations and compilers. Since it is still under development, QIR is described in natural language and lacks a formal definition, leading to ambiguity in its interpretation and a lack of rigor in implementing quantum functions. In this paper, we provide formal definitions for the data types and instruction sets of QIR, aiming to provide correctness and safety guarantees for operations and intermediate code conversions in QIR. To validate our design, we show some samples of unsafe QIR code where errors can be detected by our formal approach.
\end{abstract}

\section{Introduction}

Developing a Noisy Intermediate-Scale Quantum Computer (NISQ)~\cite{preskill2018quantum} offers new opportunities for research and development of quantum software. With the increasing development of quantum hardware, quantum program processors (QPUs) are expected to complement further and accelerate existing classical scientific computation workflows, called heterogeneous quantum-classical computation. Various programming languages such as \texttt{Qiskit}~\cite{aleksandrowicz2019qiskit}, \texttt{Cirq}~\cite{cirq2018google}, \texttt{Q\#}~\cite{svore2018q}, \texttt{Quipper}~\cite{green2013quipper}, and \texttt{ProjectQ}~\cite{projectq2017projectq} have been developed to implement such a computational model. Like classical programs, compiled quantum programming languages, such as Q\#, require their quantum programs to be compiled by a compiler into an intermediate representation (IR), which is subsequently optimized and transformed for efficient execution on a designated platform. Since IR has platform-independent features, its optimizer and executable generator can be reused under multiple source languages and generate the corresponding executable code according to the target execution platform. It is often necessary to develop new IRs or extend existing ones to adapt the intermediate representation to quantum properties. Some of the intermediate representations used in quantum programming languages include MLIR~\cite{mlir,MLIR-Quantum}, SQIR~\cite{SQIR}, and OpenQASM~\cite{OpenQASM}.

Quantum Intermediate Representation (QIR)~\cite{QIR} is a new intermediate representation of quantum programs developed by Microsoft, based on the popular open-source LLVM intermediate language~\cite{LLVM:CGO04}. QIR enables the representation of quantum computing workflows and tasks by specifying a set of rules for representing quantum structures in LLVM without any extensions or modifications. Its goal is to provide a unified and common interface to multiple quantum programming languages and quantum computing platforms, thus facilitating the development of general quantum compilation tools that can be reused in the compiler mechanism. 

\begin{figure}[h]
    \centering
    \includegraphics[width=0.8\textwidth]
    {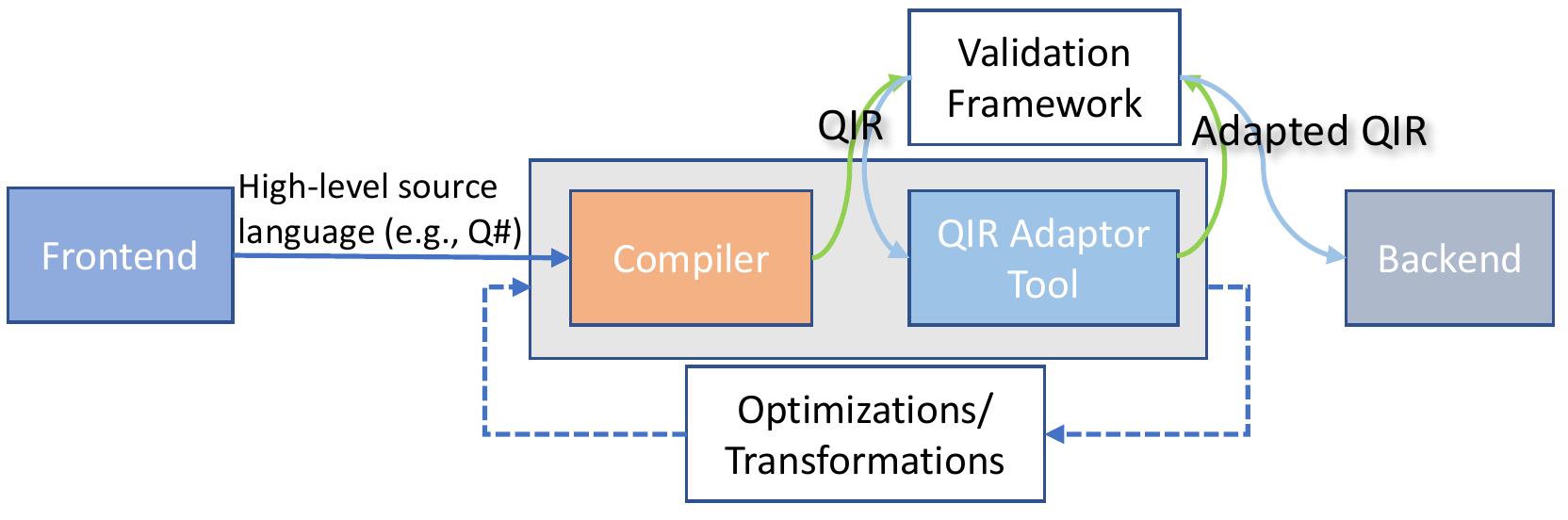}
    \caption{The QIR compiler architecture. }
    \label{fig:QIR_structure}
    \vspace*{-2mm}
\end{figure}

The QIR-based compiler processes the high-level source quantum language by converting it to QIR and handing it over to the target backend for execution (see Figure \ref{fig:QIR_structure}). Since QIR is designed and implemented based on LLVM, applying the classical LLVM optimization methods to transform and optimize QIR code is possible. However, since QIR is still in the early stage of development, it lacks clear formal semantics. Thus, its execution and the correctness of the optimization and conversion are difficult to rigorously proven, making it difficult to guarantee the correct operations of quantum programs. \textcolor{black}{In our work, we hope to design a formal method for QIR and develop a framework for verification based on this method so that QIR code can be tested for \textcolor{black}{safety} by this verification framework before it actually executed.}

In our work, we focus on the core functions of QIR, which play a crucial role in the correct operation of QIR programs. Therefore, in this paper, we prioritize formalizing the syntax and semantics of these core functions so that we can detect possible errors in the execution of QIR's core functions and guarantee that QIR's essential functions can be executed properly. Our contributions can be summarized as follows: 

\begin{itemize}
    \item We formalize the syntax of QIR based on the work of Zhao {\it et al.}~\cite{vellvm}. Specifically, we have adapted its abstract syntax to remove LLVM directives and types not used in QIR and augment QIR-specific data types. 
    \item We design the semantics of important instructions in QIR, such as allocation and release qubits, gate operations, and measurement. These operations constitute the implementation of the most basic quantum program functions, and formalization based on them can allow us to capture unsafe parts of the code. 
    \item We design a management model for qubits through which, together with our formal methods, the unsafe parts of the QIR code can be captured (e.g., qubits cloning and the use of released qubits).
    \item We validate the effectiveness of our formal approach by applying it to real cases of unsafe QIR code.
\end{itemize}

The rest of the paper is organized as follows. We introduce the background information of LLVM Intermediate Representation (LLVM IR) and QIR in Section \ref{sec:background}. We present our formalization of the syntax and semantics of QIR in Section \ref{sec:methodology}. The validation of our formal method with real-world examples is presented in Section \ref{sec:result}. We discuss related work in Sections \ref{sec:related-work}, and the conclusion is given in Section~\ref{sec:conclusion}.

\section{Background}
\label{sec:background}
This section briefly introduces LLVM IR, the basis of QIR, and QIR itself.

\subsection{Basic Concepts}

Unlike classical programs that use classical bits to store information, in quantum programs, quantum bits ("qubits") are used as the medium for storing data. As compared to a classical bit that can only be in either $0$ or $1$ state, a qubit can be in both $0$ and $1$ states, and it is called a quantum \texttt{superposition} state. The superposition state of a qubit can be represented by $\ket{\psi} = \alpha\ket{0} + \beta\ket{1}$, where $\ket{}$ is called the Dirac symbol, and $|\alpha|^2 + |\beta|^2 = 1$. We cannot directly observe the state of a qubit, which is in the superposition state; instead, we must measure it and obtain a $0$-state with probability $|\alpha|^2$ or a $1$-state with probability $|\beta|^2$.

In quantum computing, we usually use quantum logic gates to control the state of the qubits. Some basic gate operations include: 
\begin{itemize}
    \item \textbf{X gate (NOT gate)}: When a quantum undergoes the X-gate operation, its state changes from $\ket{\psi} = \alpha\ket{0} + \beta\ket{1}$ to $\ket{\psi} = \beta\ket{0} + \alpha\ket{1}$. We can use the matrix to represent this operation: 
    \begin{equation}
        X=\left[\begin{array}{ll}
0 & 1 \\
1 & 0
\end{array}\right]
    \end{equation}
    \item \textbf{Y gate and Z gate}: Similarly, for the Y and Z gates have the following expressions: 
    \begin{equation}
        Y=\left[\begin{array}{ll}
0 & -1i \\
1i & 0
\end{array}\right] \quad\&\quad Z=\left[\begin{array}{ll}
1 & 0 \\
0 & -1
\end{array}\right]
    \end{equation}
    \item \textbf{Hadamard gate (H gate)}: Hadamard gate can turn a $\ket{0}$ state or a $\ket{1}$ state into a superposition of $\ket{0}$ and $\ket{1}$ with equal probability: 
    \begin{equation}
        H=\frac{1}{\sqrt{2}}\left[\begin{array}{cc}
1 & 1 \\
1 & -1
\end{array}\right]
    \end{equation}
    \item \textbf{Controlled gate}: Unlike the single-qubit gate that can only accept one qubit as input, as described above, the controlled gate can accept multiple qubits as input. Specifically, for qubits input to the controlled gate, there are control qubits and target qubits, and the state of the target qubit is changed only when the states of both control qubits are $\ket{1}$. For example, for the CNOT gate (Controlled NOT gate), the logical expression is: 
    \begin{equation}
        \ket{A, B} \rightarrow \ket{A, B\oplus A} 
    \end{equation}
    Similarly, the Toffoli gate, also known as the CCNOT gate, possesses two control qubits that together control the flip of the target qubit with the expression: 
    \begin{equation}
        \ket{A, B, C} \rightarrow \ket{A, B, C\oplus AB}
    \end{equation}
\end{itemize}
All the above quantum logic gates are reversible, so no information loss occurs during quantum computing. By applying gate operations to qubits, quantum circuits can be composed, and thus quantum algorithms can be implemented. It is the model widely used by quantum programs nowadays. One other important operation is measurement. A classical bit in a 0 or 1 state can be obtained by performing a measurement operation on a single qubit. More content related to quantum computing can be found in the book by Nielsen and Chuang~\cite{nielsen2002quantum}.

\subsection{LLVM IR}
The LLVM IR~\cite{LLVM:CGO04} is a high-level static single assignment form (SSA) language used by the LLVM compiler infrastructure as the intermediate representation for source code. It is a representation of the source code that is independent of the source languages and target platforms. It is currently widely used and there are many conversion and optimization methods available around it. Compilers can perform various optimizations on the code of a program based on this, including common subexpression elimination, dead code elimination, and loop unrolling. These optimizations can improve the performance of the generated machine code, making it run faster and more efficiently.

LLVM IR is a strongly typed language, meaning each value has a specific type, such as an integer or floating point value. Such a design helps prevent type errors and allows the compiler to generate more efficient code. \textcolor{black}{Similar to RISC instructions,} LLVM IR is designed as a triple-address form. Combined with its strong typing, it allows for both compilation and optimization based on individual compilation units and precise global optimization performed at link time using type information. LLVM IR also includes support for control flow, loops, and other common language constructs, which makes it possible to represent a wide range of programs. 



\subsection{Quantum Intermediate Representation (QIR)}

Microsoft has developed a new quantum intermediate representation called Quantum Intermediate Representation (QIR), which is based on the widely used open-source LLVM IR. To promote QIR and provide solutions for efficient use of different quantum processors, Microsoft has established the QIR Alliance~\cite{QIR-Alliance}. QIR avoids the need for creating new compilers for different quantum programming languages by leveraging the existing classical infrastructure of LLVM IR, enabling support for the necessary functions of quantum programs without requiring additional extensions or modifications. Specifically, QIR preserves the LLVM primitives such as \textbf{call}, \textbf{bitcast}, \textbf{getelementptr}, and other classical data types such as integer \textbf{i}$sz$ and double \textbf{Double}. In order to implement quantum operations, QIR adds two data types to LLVM IR, \textbf{\%Qubit} and \textbf{\%Result}, which represent quantum registers and the results of measurements, respectively. There are also two data structures, \textbf{\%Array} and \textbf{\%Tuple}, which represent arrays of the same type of data and user-defined structures composed of arrays of multiple types, respectively. All the above four data types are opaque to QIR. To make it easier to express the operations of qubits, QIR adds new data types, such as \textbf{\%Pauli}, which refers to the four Pauli matrices in quantum mechanics.

In addition to the data types, QIR also declares quantum-related runtime functions and functions that operate on qubits (e.g., quantum gate operations and measurement operations). These functions are not implemented in the QIR and need to be defined and implemented externally. Such a design facilitates the adaptation of the QIR to different hardware backends, thus making the QIR independent of the hardware backend. QIR's official documentation~\cite{qir-spec} contains definitions for its instructions, and we will provide a brief explanation of the QIR instructions used in our work. Since the instructions for qubits declared in QIR have the same prefix \emph{@\_\_quantum\_\_rt\_\_} or \emph{@\_\_quantum\_\_qis\_\_}, in our work, we have adopted an omitted notation for them to save space (e.g., the \emph{@\_\_quantum\_\_rt\_\_qubit\_allocate} is omitted and noted as \textit{qubit\_allocate}, the \textit{@\_\_quantum\_\_qis\_\_gateop\_\_body} is omitted and noted as \textit{gateop\_\_body}). 

Figure \ref{fig:QIR_SAMPLE} shows a sample QIR code, which briefly describes the design of QIR. Lines 1 to 3 of the code mark the opaque data types. 
In lines 5 through 12, the code implements the function of allocating a qubit register and measuring it after executing the H gate
. In line 7, the \textit{qubit\_allocate} function is called, which allocates a qubit register. The functions that perform operations on the $Qubit$ are all declared as \textit{gateop\_\_body}, where \textit{gateop} is the specific operation to be performed, e.g., \textit{h} in line 8 refers to the \text{H} gate
. After executing the H gate operation in line 8, the measurement operation in line 9 is explicitly defined in lines 14 through 32. After completing the measurement, the qubit is released with \textit{qubit\_release} in line 10, and the measurement result is returned in line 11.

\begin{figure}[h]
\begin{center}
\begin{CodeOut}
\scriptsize{
\begin{alltt}

1   \%Result = type opaque
2   \%Qubit = type opaque
3   \%Array = type opaque
4   
5   define internal \%Result* @Sample__SampleQ__body() \{
6   entry:
7     \%q = call \%Qubit* @__quantum__rt__qubit_allocate()
8     call void @__quantum__qis__h__body(\%Qubit* \%q)
9     \%0 = call \%Result* @Microsoft__Quantum__Intrinsic__M__body(\%Qubit* \%q)
10    call void @__quantum__rt__qubit_release(\%Qubit* \%q)
11    ret \%Result* \%0
12  \}
13  
14  define internal \%Result* @Microsoft__Quantum__Intrinsic__M__body(\%Qubit* \%qubit) \{
15  entry:
\iffalse
16    \%bases = call \%Array* @__quantum__rt__array_create_1d(i32 1, i64 1)
17    \%0 = call i8* @__quantum__rt__array_get_element_ptr_1d(\%Array* \%bases, i64 0)
18    \%1 = bitcast i8* \%0 to i2*
19    store i2 -2, i2* \%1, align 1
20    call void @__quantum__rt__array_update_alias_count(\%Array* \%bases, i32 1)
21    \%qubits = call \%Array* @__quantum__rt__array_create_1d(i32 8, i64 1)
22    \%2 = call i8* @__quantum__rt__array_get_element_ptr_1d(\%Array* \%qubits, i64 0)
23    \%3 = bitcast i8* \%2 to \%Qubit**
24    store \%Qubit* \%qubit, \%Qubit** \%3, align 8
25    call void @__quantum__rt__array_update_alias_count(\%Array* \%qubits, i32 1)
26    \%4 = call \%Result* @__quantum__qis__measure__body(\%Array* \%bases, \%Array* \%qubits)
27    call void @__quantum__rt__array_update_alias_count(\%Array* \%bases, i32 -1)
28    call void @__quantum__rt__array_update_alias_count(\%Array* \%qubits, i32 -1)
29    call void @__quantum__rt__array_update_reference_count(\%Array* \%bases, i32 -1)
30    call void @__quantum__rt__array_update_reference_count(\%Array* \%qubits, i32 -1)
\fi
    ...
31    ret \%Result* %4
32  \}
33
34    declare \%Qubit* @__quantum__rt__qubit_allocate() 
35    declare void @__quantum__rt__qubit_release(\%Qubit*) 
36    declare void @__quantum__qis__h__body(\%Qubit*) 
37    declare void @__quantum__rt__array_update_alias_count(\%Array*, i32) 
38    declare \%Array* @__quantum__rt__array_create_1d(i32, i64) 
39    declare i8* @__quantum__rt__array_get_element_ptr_1d(\%Array*, i64) 
40    declare \%Result* @__quantum__qis__measure__body(\%Array*, \%Array*) 
41    declare void @__quantum__rt__array_update_reference_count(\%Array*, i32) 
    
\end{alltt}
    }
    \end{CodeOut}
    \caption{Example of QIR.}
    \label{fig:QIR_SAMPLE}
\end{center}
\end{figure}

\section{Methodology}
\label{sec:methodology}

In this section, we formalize the syntax and semantics of QIR to support the verification of unsafe code. Since QIR is designed based on LLVM IR, this paper does not consider the formalization of the classical LLVM IR part, which can be done using methods~\cite{vellvm, k-llvm}. Our formalization of QIP is based on version 0.1 of QIR.

\subsection{Syntax of QIR}

We first present our formalization of the syntax of QIR in Figure~\ref{fig:QIR_syntax}. Our abstract syntax extends on the work of Zhou {\it et al.}~\cite{vellvm}, which provides a relatively complete formalization of LLVM IR. We mainly add to it data types that are specific to QIR, such as \textbf{\%Qubit}, \textbf{\%Result}, \textbf{\%Array}, etc. Since the QIR instruction is called with LLVM's \textbf{call} instruction, its abstract semantics can be attributed to $id$, so instructions about quantum gate operations, measurement operations, etc., cannot be reflected in the abstract syntax.

\begin{figure}[h]
    \centering
    \begin{math}
    \small
        \begin{array}{l}
        \text{Types\quad\quad\ $typ$ ::= \textbf{i}\textit{sz} $|$ \textbf{double} $|$ \textit{typ}* $|$ [$sz\times typ$] $|$ $typ$ $\overline{typ_j}^j$ $|$ \{$\overline{typ_j}^j$\} $|$ \textit{id} $|$ \textcolor{red}{\textbf{\%Pauli}} $|$ \textcolor{red}{\textbf{\%Range}} $|$ \textcolor{red}{\textbf{\%Result}} }\\
        \text{\quad\quad\quad\quad\quad\quad\ \ \  $|$ \textcolor{red}{\textbf{\%Qubit}} $|$ \textcolor{red}{\textbf{\%Array}} $|$ \textcolor{red}{\textbf{\%Tuple}}}\\
            \text{Products\ \ \textit{prod} ::= \textbf{define} $typ$ $id$ ($\overline{arg}$) \{$\overline{b}$\} $|$ \textbf{declare} $typ$  $id$ ($\overline{arg}$)}\\
            \text{Values\quad\ \ \ \ $val$ ::= $id$ $|$ $cnst$} \\
            \text{Constants \textit{cnst} ::= \textbf{i}\textit{sz} \textit{Int} $|$ \textbf{double} \textit{Double} $|$ 
 \textbf{void} $|$ $typ^* id$ $|$ $typ$ [$\overline{cnst_j}^j$] $|$ \{$\overline{cnst_j}^j$\} $|$ \textbf{extractvalue} $cnst$ $\overline{cnst_j}^j$ }\\
            \text{\quad\quad\quad\quad\quad\quad\ \  $|$ \textbf{getelementptr} $cnst$ $\overline{cnst_j}^j$ $|$ \textbf{bitcast} \textit{cnst} \textbf{to} \textit{typ} $|$ \textcolor{red}{\textbf{\%Range} \textit{Int} \textit{Int} \textit{Int}} $|$ \textcolor{red}{\textbf{\%Pauli} $Int$}}\\
            \text{Blocks\ \ \quad\quad b ::= $l\overline{\phi}\overline{c}tmn$} \\
            \text{Tmns \quad\ \ \ $tmn$ ::= \textbf{ret} $typ$ $val$ $|$ \textbf{ret void} $|$ \textbf{br} $val$ $l_1$ $l_2$ $|$ \textbf{br} $l$ }\\
            \text{Commands \ \ \textit{c} ::= \textit{id} = \textbf{bitcast} $typ_1 val$ \textbf{to} $typ_2$ $|$ \textbf{store} \textit{typ} $val_1$ $val_2$ \textit{align} $|$ \textit{option id} = \textbf{call} $typ_0$ $val_0$ $\overline{param}$ }\\
            \text{\quad\quad\quad\quad\quad\quad\ \  $|$ \ \textit{id} = \textbf{getelementptr} ($typ*$) $val$ $\overline{val_j}^j$ $|$ \textit{id} = \textbf{load} ($typ*$) $val_1$ $align$}\\
            \text{\quad\quad\quad\quad\quad\quad\ \  $|$ \ $id$ = \textbf{icmp} $cond$ $typ$ $val_1$ $val_2$ $|$ $id$ = \textbf{alloca} $typ$ $val$ $align$}\\
        \end{array}
    \end{math}
    \caption{The abstract syntax of QIR. The red font shows what is added to the abstract semantics of QIR compared to LLVM IR.}
    \label{fig:QIR_syntax}
\end{figure}

\subsection{Semantics of QIR}

We next present our formalization of the semantics of QIR. Since QIR is designed to treat instructions as declarations of quantum runtime functions, the type is explicitly specified for each instruction input and output. Such a design avoids the nondeterministic semantics found in classical LLVM IR, thus liberating our efforts from cumbersome type checking and allowing us to focus the formalization on checking the correctness of quantum operations. Since the back-end hardware is responsible for implementing quantum operations in QIR, formalizing this process in QIR is not necessary. The guarantee of correct quantum operations should lie with the back-end hardware. Therefore, we can assume that the quantum operations are correct and further improve the security of the QIR code.

\subsubsection{Allocating and Releasing Qubits}
\label{Allocate}

QIR manipulates the corresponding qubits in the back-end registers by feeding $\textbf{\%Qubit}^*$ or $\textbf{\%Array}^*$ composed of them to the corresponding functions, without having direct access to $\textbf{\%Qubit}$ or $\textbf{\%Array}$ information (since they are opaque to QIR as well as LLVM). It makes it difficult for QIR to manage the state of qubits, and the compiler cannot know whether they have been released, thus raising a potential security risk. In this section, we formalize the semantics from the creation of a single qubit as well as an array of qubits and propose a set of management schemes for qubits and qubit arrays so that the possible problems of qubits cloning and the use of released qubits in QIR can be effectively captured.

\paragraph{Static creation.}
In QIR, the allocation of qubits includes both static and dynamic methods. For qubits whose identifiers are already determinable at compile time, static qubit values can be generated using LLVM's \textbf{inttoptr} instruction. The following example is given in the specification document of QIR: 

\begin{center}
    \text{\%qubit3 = \textbf{{inttoptr}} \textbf{i}32 3 \textbf{{to}} $\textbf{\%Qubit}^*$}
    \vspace*{-2mm}
\end{center}

This code refers to a qubit3 device in the real equipment (or simulator) by converting the 3 of the \textbf{i}32 data type to the $\textbf{\%Qubit}^*$ type. \textcolor{black}{Since the instructions for static creation of qubits use only the classical LLVM IR, the formal semantics are not elaborated in our work. }

\begin{figure}
    \centering
    \includegraphics[width=0.95\textwidth]{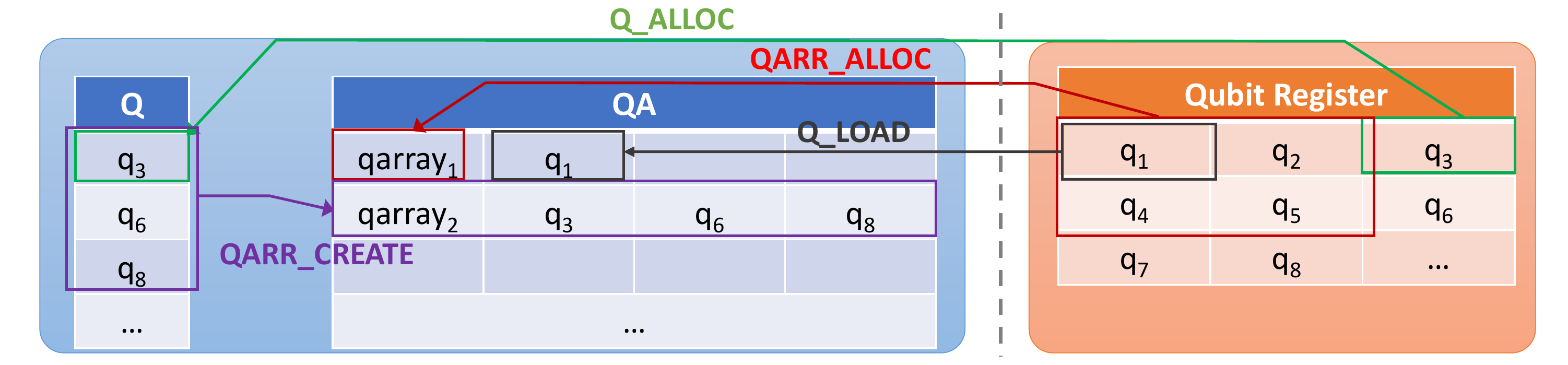}
    \caption{A management model for qubits and qubits arrays in QIR.}
    \label{fig:QandQAmanagement}
    \vspace*{-2mm}
\end{figure}

\paragraph{Dynamic management.} 
Dynamic qubits are allocated and released by the quantum runtime method. To store the allocated qubits and qubit arrays in the QIR code, we introduce two data structures in the framework, a one-dimensional array \textbf{\texttt{Q}} and a two-dimensional array \textbf{\texttt{QA}}, respectively. Our management model is presented in Figure \ref{fig:QandQAmanagement}, where \textbf{\texttt{Q}} stores the $\textbf{\%Qubit}^*$ pointers for each allocated qubit. The first column of \textbf{\texttt{QA}} stores the $\textbf{\%Array}^*$ pointer for the qubit array, while the corresponding row stores the pointer $\textbf{\%Qubit}^*$ for each qubit in the array. Some of the methods used to manage these two data structures are listed in Table \ref{tab: QQAmanagement}.

\begin{table}[h]
\centering
\caption{Some of the methods used to manage qubits and qubit arrays in our validation framework}
\begin{tabular}{|l|l|}
\hline
\makecell[c]{Method}                                 & \makecell[c]{Description}                                                             \\ \hline
\textbf{appqlist}(\textbf{\texttt{Q}}, $q$)  & Append $q$ to \textbf{\texttt{Q}}.                                              \\ \hline
\textbf{appqarrlist}(\textbf{\texttt{QA}}, $qarray$) & Append $qarray$ to \textbf{\texttt{QA}}.                                              \\ \hline
\textbf{checkq}(\textbf{\texttt{Q}}, $q$) & Check if $q$ is in \textbf{\texttt{Q}}. If true, return 1; otherwise, return 0.    \\ \hline
\textbf{checkqarrlist}(\textbf{\texttt{QA}}, $qarray$) & Check if $qarray$ is in \textbf{\texttt{QA}}. If true, return 1; otherwise, return 0.    \\ \hline
\textbf{delq}(\textbf{\texttt{Q}}, $q$)         & Remove $q$ from \textbf{\texttt{Q}}.                                            \\ \hline
\textbf{delqarr}(\textbf{\texttt{QA}}, $qarray$)         & Remove the row of $qarray$ from \textbf{\texttt{QA}}.                                            \\ \hline
\textbf{appqarr}(\textbf{\texttt{QA}}, $qarray$, $q$)                 & Append $q$ to $qarray$ in \textbf{\texttt{QA}}. Skip if $q$ already exists in $qarray$.                                              \\ \hline
\textbf{checkqarr}(\textbf{\texttt{QA}}, $qarray$, $q$)               & Check if $q$ is in $qarray$. If true, return 1; otherwise, return 0.           \\ \hline
\textbf{findqarr}(\textbf{\texttt{QA}}, $q$)                        & Returns the array where $q$ is located, or 0 if the array does not exist. \\ \hline
\end{tabular}
\label{tab: QQAmanagement}
\end{table}

Our formal method is concerned with two possible \textcolor{black}{safety} issues in QIR code: the \textbf{cloning operation of qubits} and the \textbf{usage of released qubits}. These two kinds of unsafe codes can be effectively captured using our proposed management model. Equations \ref{QALLOC} to \ref{QARRDEALLOC} show the operational semantics of allocation and release of qubits and qubit arrays, respectively, where $qubit\_allocate$, $qubit\_allocate\_array$, $qubit\_release$, and $qubit\_release\_array$ are the instructions provided in QIR. Specifically, after performing the allocation of qubits and qubit arrays, their return values ($\textbf{\%Qubit}^*$ and $\textbf{\%Array}^*$) are stored in \textbf{\texttt{Q}} and \textbf{\texttt{QA}} (See Figure \ref{fig:QandQAmanagement}). After releasing the qubit and qubit arrays, the corresponding values are removed from \textbf{\texttt{Q}} and \textbf{\texttt{QA}}. \textbf{To avoid using released qubits and qubit arrays}, we add the process of checking whether the qubits and qubit arrays are in \textbf{\texttt{Q}} and \textbf{\texttt{QA}} (use \textbf{checkq} and \textbf{checkqarrlist}) in all the rules in addition to \hyperref[QALLOC]{Q\_ALLOC} and \hyperref[QARRALLOC]{QARR\_ALLOC}. 

In particular, when releasing a single qubit (rule \hyperref[QDEALLOC]{Q\_DEALLOC}), we have added the process of checking whether the qubit belongs to a qubit array (with \textbf{findqarr}) because releasing a single qubit in a qubit array directly can lead to errors when releasing the qubit array later.

\begin{equation}
\footnotesize
    \begin{aligned}
        \frac{\begin{aligned}
            q=qubit\_allocate()\ \textbf{appqlist}(\textbf{\texttt{Q}}, q)=\textbf{\texttt{Q}}'
        \end{aligned}}{\textbf{\texttt{Q}} \vdash \textbf{\texttt{Q}},id=\textbf{call \%Qubit}^* qubit\_allocate()\Rightarrow \textbf{\texttt{Q}}', id\leftarrow q}\ \text{Q\_ALLOC}
    \end{aligned}
    \vspace*{-2mm}
    \label{QALLOC}
\end{equation}

\begin{equation}
\footnotesize
    \begin{aligned}
        \frac{\begin{aligned}
            qarray=qubit\_allocate\_array(n)\ \textbf{appqarrlist}(\textbf{\texttt{QA}}, qarray)=\textbf{\texttt{QA}}'
        \end{aligned}}{\textbf{\texttt{QA}}, n\vdash \textbf{\texttt{QA}}, id=\textbf{call \%Array}^* qubit\_allocate\_array(n)\Rightarrow \textbf{\texttt{QA}}', n, id\leftarrow qarray}\ \text{QARR\_ALLOC}
    \end{aligned}
    \vspace*{-4mm}
    \label{QARRALLOC}
\end{equation}

\begin{equation}
\footnotesize
\begin{aligned}
    \frac{\begin{aligned}
        &\text{if not}\ qarray=\textbf{findqarr}(\textbf{\texttt{QA}}, q)\ \text{and}\ \textbf{checkq}(\textbf{\texttt{Q}}, q)\ \text{then}\ qubit\_release(q)\ \textbf{delq}(\textbf{\texttt{Q}},q)=\textbf{\texttt{Q}}'\ \text{else}\ abort
    \end{aligned}}{\textbf{\texttt{Q}}, \textbf{\texttt{QA}}, q \vdash \textbf{\texttt{Q}}, \textbf{\texttt{QA}}, \textbf{call void}\ qubit\_release(q) \Rightarrow \textbf{\texttt{Q}}', \textbf{\texttt{QA}}, q}\ \text{Q\_DEALLOC}
\end{aligned}
\label{QDEALLOC}
\end{equation}

\begin{equation}
\footnotesize
    \begin{aligned}
        \frac{\begin{aligned}
            &\text{if}\ \textbf{checkqarrlist}(\textbf{\texttt{QA}}, qarray)\ \text{then}\ qubit\_release\_array(qarray)\ \textbf{delqarr}(\textbf{\texttt{QA}},qarray)=\textbf{\texttt{QA}}'\ \text{else}\ abort
        \end{aligned}}{\textbf{\texttt{QA}}, qarray\vdash \textbf{\texttt{QA}}, \textbf{call void}\ qubit\_release\_array(qarray)\Rightarrow \textbf{\texttt{QA}}', qarray}\ \text{QARR\_DEALLOC}
    \end{aligned}
    \label{QARRDEALLOC}
\end{equation}

\subsubsection{Manipulating Qubits in Qubit Arrays and Forming Qubit Arrays with Allocated Qubits}

In addition to directly allocating qubits and qubit arrays with the methods of QIR, it is possible to load a single qubit from a qubit array or combine multiple allocated qubits into a new qubit array.  We summarize these two operations in the functions $load\_qubit$ and $create\_qubit\_array$, respectively: 
\\
\\
\begin{math}
\small
    \begin{aligned}
        (id=load\_qubit(qarray, n))=\{&id_1=\textbf{call i}sz^*\ array\_get\_element\_ptr\_1d(qarray, n);\\&id_2=\textbf{bitcast i}sz^*\ id_1\ \textbf{to}\ \textbf{\%Qubit}^{**};\\&id=(id_3=\textbf{load \%Qubit}^*, \textbf{\%Qubit}^{**}\ id_2, align)\}
    \end{aligned}
\end{math}\\
\\
\begin{math}
\small
    \begin{aligned}
        (id=&create\_qubit\_array(q_1, q_2, ...,q_n))=\{\\&id=(id_1=\textbf{call \%Array}^* array\_create\_1d(sz, n));\\
        &\overline{id_{2i}=(\textbf{call i}sz^*\ array\_get\_element\_ptr\_1d(qarray, i));id_{3i}=\textbf{bitcast i}sz^*\ id_{2i}\ \textbf{to}\ \textbf{\%Qubit}^{**};}^{i=1...n}\\
        &\overline{\textbf{store \%Qubit}^*\ q_i, \textbf{\%Qubit}^{**}\ id_{3i}, align;}^{i=1...n}\}
    \end{aligned}
\end{math}\\

In the above two functions, $array\_create\_1d$ and $array\_get\_element\_ptr\_1d$ are instructions in QIR. $array\_create\_1d$ returns an array pointer $\textbf{\%Array}^*$ of size $sz$ with length $n$ by inputting byte size $sz$ of array elements and array length $n$. $array\_get\_element\_ptr\_1d$ can obtain the $\textbf{i}sz^*$ pointer of the corresponding element by inputting the array $qarray$ and indices $n$ of the qubit array. Then through the \textbf{bitcast} instruction of LLVM, it can be converted into a $\textbf{\%Qubit}^{**}$ pointer and the qubit pointer $\textbf{\%Qubit}^{*}$ can be loaded or stored with \textbf{load} or \textbf{store} instruction.

\paragraph{Manipulating qubits in qubit arrays. }The rule \hyperref[QLOAD]{Q\_LOAD} defines the semantics of loading a qubit from the qubit array. Specifically, besides the necessity to check whether the qubit array has been released before executing the loaded operation, it is also needed to record the qubit in the corresponding qubit array of the list \textbf{\texttt{QA}} at the end of the execution (see Figure \ref{fig:QandQAmanagement}).

\begin{equation}
\footnotesize
    \begin{aligned}
        \frac{\begin{aligned}
            &\text{if}\ \textbf{checkqarrlist}(\textbf{\texttt{QA}}, qarray)\ \text{then}\ ptr=(id_1=\textbf{call i}sz^* array\_get\_element\_ptr\_1d(qarray, n))\\ 
            &qptr=(id_2=\textbf{bitcast i}sz^*\ ptr\ \textbf{to}\ \textbf{\%Qubit}^{**})\ q=(id_3=\textbf{load}\ \textbf{\%Qubit}^*, \textbf{\%Qubit}^{**}\ qptr, align)\\ 
            &\textbf{appqarr}(\textbf{\texttt{QA}}, qarray, q)=\textbf{\texttt{QA}}'\ \text{else}\ abort
        \end{aligned}}{\textbf{\texttt{QA}}, qarray,n\vdash \textbf{\texttt{QA}}, id=load\_qubit(qarray,n)\Rightarrow \textbf{\texttt{QA}}', qarray, n, id\leftarrow q}\ \text{Q\_LOAD}
    \end{aligned}
    \label{QLOAD}
\end{equation}

\paragraph{Forming qubit arrays with allocated qubits. }The rule \hyperref[QARRCREATE]{QARR\_CREATE} summarizes the semantics of the function that forms an array qubit from allocated qubits. We add the process of checking whether the qubit to be stored already exists in the array (\textbf{checkqarr}) before the store instruction and abort if the qubit already exists in the array. It is because there should not be two identical qubits in the same qubit array, otherwise, there will be \textbf{a risk of qubits cloning}. If this qubit array is used as control qubits, two identical qubits will act as control qubits simultaneously. While it doesn't impact the operation's outcome, it violates the non-cloning principle of quantum computing. Its management for \textbf{\texttt{Q}} and \textbf{\texttt{QA}} is shown in Figure \ref{fig:QandQAmanagement}. Specifically, rule \hyperref[QARRCREATE]{QARR\_CREATE} adds the new array pointer to the \textbf{\texttt{QA}} and adds the qubits that make up the array to the corresponding row.

\begin{equation}
\footnotesize
    \begin{aligned}
        \frac{\begin{aligned}
            &\text{if}\ \overline{\textbf{checkq}(\textbf{\texttt{Q}}, q_i)\ \text{or}\ qarray_{temp}=\textbf{findqarr}(\textbf{\texttt{QA}}, q_i)}^{i=1...n}\ \text{then}\\\
            &qarray=(id_1=\textbf{call \%Array}^* array\_create\_1d(sz, n))\ \textbf{appqarrlist}(\textbf{\texttt{QA}}, qarray)\ \\
            &\overline{ptr_i=(id_{2i}=(\textbf{call i}sz^*\ array\_get\_element\_ptr\_1d(qarray, i)))\ qptr_i=(id_{3i}=\textbf{bitcast i}sz^*\ ptr_i\ \textbf{to}\ \textbf{\%Qubit}^{**})}^{i=1...n}\\
            &\overline{\text{if not}\ \textbf{checkqarr}(\textbf{\texttt{QA}}, qarray, q_i)\ \text{then}\ \textbf{store \%Qubit}^*\ q_i, \textbf{\%Qubit}^{**}\ qptr_i, align\ \text{else}\ abort}^{i=1...n}\\
            &\overline{\textbf{appqarr}(\textbf{\texttt{QA}}, qarray, q_i)}^{1=1...n}=\textbf{\texttt{QA}}'\ \text{else}\ abort
        \end{aligned}}{\textbf{\texttt{Q}}, \textbf{\texttt{QA}}, (q_1, q_2, ...,q_n) \vdash \textbf{\texttt{Q}}, \textbf{\texttt{QA}}, id=create\_qubit\_array(q_1, q_2, ...,q_n)\Rightarrow \textbf{\texttt{Q}}, \textbf{\texttt{QA}}', (q_1, q_2, ...,q_n), id\leftarrow qarray}\ \text{QARR\_CREATE}
    \end{aligned}
    \label{QARRCREATE}
\end{equation}

\subsubsection{Gate Operations}
In QIR, gate operations can be divided into two main categories: single-qubit gate operations and controlled gate operations formed by adding control qubits to single-qubit gate operations. We will formalize the semantics of these two types of gate operations separately. In QIR, the gate operations of qubits do not change in our management model for \textbf{\texttt{Q}} and \textbf{\texttt{QA}} since they do not involve loading and storing of data as well as allocation and release of qubits. The rules \hyperref[SGOP]{SG\_OP} and \hyperref[CGOP]{CG\_OP} show our formalization of the semantics of single-qubit gates as well as controlled gates, respectively, where $gateop\_\_body$ and $gateop\_\_ctl$ are instructions for gate operations in QIR, and $gateop$ is a single-qubit gate such as $x$,$Rx$,$h$.

\begin{equation}
\footnotesize
    \begin{aligned}
        \frac{\begin{aligned}
            &\text{if}\ \textbf{checkq}(\textbf{\texttt{Q}}, q)\ \text{or}\ qarray_{temp}=\textbf{findqarr}(\textbf{\texttt{QA}}, q)\ \text{then}\ gateop\_\_body(option\ d, q)\ \text{else}\ abort
        \end{aligned}}{\textbf{\texttt{Q}}, \textbf{\texttt{QA}}, option\ d, q\vdash \textbf{\texttt{Q}}, \textbf{\texttt{QA}}, \textbf{call void}\ gateop\_\_body(option\ d, q)\ \Rightarrow \textbf{\texttt{Q}}, \textbf{\texttt{QA}}, option\ d, q}\ \text{SG\_OP}
    \end{aligned}
    \label{SGOP}
\end{equation}

\begin{equation}
\footnotesize
    \begin{aligned}
        \frac{\begin{aligned}
            &\text{if}\ (\textbf{checkq}(\textbf{\texttt{Q}}, q)\ \ \text{or}\ qarray_{temp}=\textbf{findqarr}(\textbf{\texttt{QA}}, q))\ \text{and}\ \textbf{checkqarr}(\textbf{\texttt{QA}}, qarray)\ \\ &\text{and not}\ \textbf{checkqarr}(\textbf{\texttt{QA}}, qarray, q)\ \text{then}\ gateop\_\_ctl(qarray,(option\ d,q))\ \text{else}\ abort
        \end{aligned}}{\textbf{\texttt{Q}}, \textbf{\texttt{QA}}, option\ d, q\vdash\textbf{\texttt{Q}}, \textbf{\texttt{QA}}, \textbf{call void}\ gateop\_\_ctl(qarray,(option\ d,q))\Rightarrow \textbf{\texttt{Q}}, \textbf{\texttt{QA}}, option\ d, q}\ \text{CG\_OP}
    \end{aligned}
    \label{CGOP}
\end{equation}

\vspace*{3mm}
In QIR, the control qubits are entered into the instruction in the form of a qubit array. So similar to the rule \hyperref[QARRCREATE]{QARR\_CREATE}, for \hyperref[CGOP]{CG\_OP}, we also add a check for whether the target qubit exists in the control qubit array (with \textbf{checkqarr}) to avoid the problem of qubit cloning. 


\label{GateOperation}

\subsubsection{Measurement}
For the measurement operation, the measurement result can be returned in QIR by entering an array of Pauli values and an array of qubits. The semantics of the measurement operation is: 

\begin{equation}
\footnotesize
    \begin{aligned}
        \frac{\begin{aligned}
            \text{if}\ \textbf{checkqarr}(\textbf{\texttt{QA}},qarray)\ \text{then}\ result=measure\_\_body(Pauliarr, qarray)\ \text{else}\ abort
        \end{aligned}}{\textbf{\texttt{QA}},Pauliarr, qarray\vdash id=\textbf{call \%Result}^*\ measure\_\_body(Pauliarr, qarray)\Rightarrow \textbf{\texttt{QA}},Pauliarr, qarray, id\leftarrow result}\ \text{MEASURE}
    \end{aligned}
    \vspace*{-2mm}
\end{equation}

\section{Verification on Unsafe Code of QIR}
\label{sec:result}

This section applies our method to verify some insecure QIR codes. Since QIR is still not officially in use, the most prominent way to generate QIR code at present is to use the Q\# compiler provided by Microsoft. This provides us with the inspiration to collect verification samples. If an unsafe Q\# program passes the compiler but fails at runtime, we can generate the corresponding QIR code from it for verification. Figure \ref{fig:Unsafe_qsharp_SAMPLE} shows two samples of unsafe Q\# code referenced from~\cite{qsharpalgo}, where in Figure \ref{fig:Unsafe_qsharp_SAMPLE}a the qubit returned is released at the end of the execution of the function \textit{NewQubit}, and in Figure \ref{fig:Unsafe_qsharp_SAMPLE}b the three qubits called by \textit{CCNOT} are the same qubit, violating the no-cloning theorem. In the next subsections, we will present the corresponding QIR code and apply the semantics we designed to check for faults.

\begin{figure*}[h]
\centering
\subfigbottomskip=2pt
\subfigcapskip=5pt
\subfigure[Return a released qubit.]{
$\footnotesize{
\begin{array}{l}
    \texttt{\@EntryPoint()}\\
    \texttt{operation Deadqubit() : Result \{}\\
        \quad\quad\texttt{let q = NewQubit();}\\
        \quad\quad\texttt{H(q);}\\
        \quad\quad\texttt{return M(q);}\\
    \texttt{\}}\\
    \\
    \texttt{operation NewQubit() : Qubit \{}\\
        \quad\quad\texttt{use q = Qubit();}\\
        \quad\quad\texttt{return q;}\\
    \texttt{\} }
\end{array}
    }
$}
\vspace*{-2mm}
\subfigure[Control qubit and target qubit are the same one.]{
$\footnotesize{
\begin{array}{l}
    $\texttt{\@EntryPoint()}$\\
    $\texttt{operation Cloning() : Result \{}$\\
    $\quad\quad\texttt{use q1 = Qubit();}$\\
    $\quad\quad\texttt{let q2 = q1;}$\\
    $\quad\quad\texttt{let q3 = q2;}$\\
    $\quad\quad\texttt{CCNOT(q1, q2, q3);}$\\
    $\quad\quad\texttt{return M(q3);}$\\
    $\texttt{\} }$   
\end{array}    
    }
$}
    \caption{Example of unsafe Q\# codes.}
    \vspace*{-5mm}
    \label{fig:Unsafe_qsharp_SAMPLE}
\end{figure*}

\subsection{Using Deallocated Qubits}
\begin{figure}
    \centering
    \includegraphics[width=0.95\textwidth]{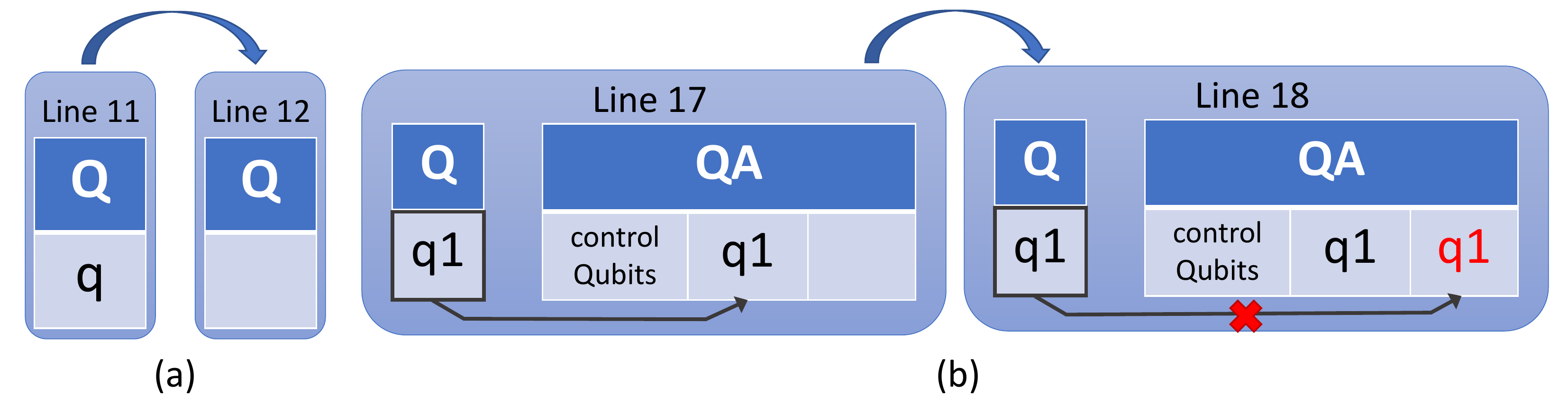}
    \caption{Example of applying the formal method to verify unsafe QIR code.}
    \label{fig:QQAexample}
\end{figure}

Figure \ref{fig:Released_qubit_QIR} shows the QIR code obtained from the Q\# code conversion from Figure \ref{fig:Unsafe_qsharp_SAMPLE}a. 
In this example, for the return value of the function $@NewQubit\_\_body$ in line 3, we can analyze it using the rules \hyperref[QALLOC]{Q\_ALLOC} and \hyperref[QDEALLOC]{Q\_DEALLOC}, so that we can know that the $\%q$ returned has been released (see Figure \ref{fig:QQAexample}a). In the execution of $h\_\_body$ in line 4, the rule \hyperref[SGOP]{SG\_OP} is applied to interrupt the program since $\%q$ does not exist in \textbf{\texttt{Q}}, thus avoiding the error of using the released qubit.

\begin{figure}[h]
\begin{center}
\begin{CodeOut}
\scriptsize{
\begin{alltt}
\quad \quad \quad 1   define internal \%Result* @Deadqubit__body() \{
\quad \quad \quad 2   entry:
\quad \quad \quad 3     \%q = call \%Qubit* @NewQubit__body()
\quad \quad \quad 4     call void @__quantum__qis__h__body(\%Qubit* \%q)
\quad \quad \quad 5     \%0 = call \%Result* @Microsoft__Quantum__Intrinsic__M__body(\%Qubit* \%q)
\quad \quad \quad 6     ret \%Result* \%0
\quad \quad \quad 7   \}
\quad \quad \quad 8
\quad \quad \quad 9   define internal \%Qubit* @NewQubit__body() \{
\quad \quad \quad 10   entry:
\quad \quad \quad 11    \%q = call \%Qubit* @__quantum__rt__qubit_allocate()
\quad \quad \quad 12    call void @__quantum__rt__qubit_release(\%Qubit* \%q)
\quad \quad \quad 13    ret \%Qubit* \%q
\quad \quad \quad 14  \} 
\quad \quad \quad     ...
\end{alltt}
    }
    \end{CodeOut}
    \caption{QIR code converted from Figure \ref{fig:Unsafe_qsharp_SAMPLE}a. To save space, we have only intercepted the minimum critical code.}
    \label{fig:Released_qubit_QIR}
\end{center}
\end{figure}

\subsection{Qubit Cloning}

Figure \ref{fig:Cloning_qubit_QIR} shows the QIR code obtained from the code conversion of the cloned quantum in Figure \ref{fig:Unsafe_qsharp_SAMPLE}b. 
In the example, for the function, $CCNOT\_\_body$ in line 4, two identical $\textbf{Qubit}^*$ $\%q1$ are entered as control qubits, and $\%q1$ itself as the target qubit. Thus for lines 12 to 18 of the code, the rule \hyperref[QARRCREATE]{QARR\_CREATE} can be applied, thus interrupting the program in line 18 due to the transfer to the array $\%\_\_controlQubits\_\_$ stores the same $\textbf{Qubit}^*$ $\%q1$ and avoiding the qubit cloning problem (see Figure \ref{fig:QQAexample}b). Other than that, for line 20 of the code, the rule \hyperref[CGOP]{CG\_OP} can be applied, and since $\%q1$, which is the target qubit, is the same as the element in the control qubit array, qubit cloning can be avoided in this step as well.

\begin{figure}
    \centering
    \begin{CodeOut}
    \scriptsize{
        \begin{alltt}
1   define internal \%Result* @Cloning__body() \{
2   entry:
3     \%q1 = call \%Qubit* @__quantum__rt__qubit_allocate()
4     call void @Microsoft__Quantum__Intrinsic__CCNOT__body(\%Qubit* \%q1, \%Qubit* \%q1, 
                                                            \%Qubit* \%q1)
    ...
8   \}
9   
10  define internal void @Microsoft__Quantum__Intrinsic__CCNOT__body(\%Qubit* \%control1, 
                                                \%Qubit* \%control2, \%Qubit* \%target) \{
11  entry:
12    \%__controlQubits__ = call \%Array* @__quantum__rt__array_create_1d(i32 8, i64 2)
13    \%0 = call i8* @__quantum__rt__array_get_element_ptr_1d(\%Array* \%__controlQubits__, i64 0)
14    \%1 = bitcast i8* \%0 to \%Qubit**
15    \%2 = call i8* @__quantum__rt__array_get_element_ptr_1d(\%Array* \%__controlQubits__, i64 1)
16    \%3 = bitcast i8* \%2 to \%Qubit**
17    store \%Qubit* \%control1, \%Qubit** \%1, align 8
18    store \%Qubit* \%control2, \%Qubit** \%3, align 8
19    call void @__quantum__rt__array_update_alias_count(\%Array* \%__controlQubits__, i32 1)
20    call void @__quantum__qis__x__ctl(\%Array* \%__controlQubits__, \%Qubit* \%target)
    ...  
24  \}   
        \end{alltt}
        }
    \end{CodeOut}
    \vspace*{-5mm}
    \caption{QIR code converted from Figure \ref{fig:Unsafe_qsharp_SAMPLE}a. To save space, we have only intercepted the minimum critical code.}
    \label{fig:Cloning_qubit_QIR}
    \vspace*{-2mm}
\end{figure}

\section{Related Work}
\label{sec:related-work}
This section discusses some related work in the areas of intermediate representations for both classical and quantum programming languages.

\subsection{LLVM IR Semantics}
As the basis of QIR, the formal approach of LLVM IR is an important reference for our work. Zhao {\it et al.}~\cite{vellvm} propose the Vellvm (verified LLVM) framework, which provides a formalization of the static semantics, memory model, and several operational semantics of LLVM IR. The framework is implemented using Coq, through which verified executable code with high confidence can be extracted directly, and the effectiveness of Vellvm was verified with the SoftBound~\cite{softbound} case. Li and Gunter~\cite{k-llvm} design K-LLVM, the complete formal LLVM IR semantics, including all LLVM IR instructions, intrinsic functions in the LLVM documentation, etc. Compared to Vellvm, which focuses on formalizing LLVM semantics as a mathematical object, K-LLVM shows possible implementations of the semantics in a virtual computer with a more direct approach. K-LLVM is implemented by $\mathbb{K}$~\cite{K}, and its validity is verified by testing against unit test programs as well as actual LLVM IR programs. However, the above work is directed toward formalizing LLVM IR and does not address quantum programming languages or intermediate representations.

\subsection{Formalized Quantum Intermediate Representation and Programming Language}

As the most relevant work to us, Hietala {\it et al.}~\cite{SQIR} present VOQC, a verified optimizer for quantum circuits. As the input to VOQC, a small quantum intermediate representation (SQIR) was developed to represent quantum circuits and support the verification of quantum circuit optimization. SQIR is well formalized, and its syntax and semantics for quantum circuits can guarantee the correctness of VOQC optimization. Our work differs from SQIR in two main ways: first, the formal object, which is QIR developed by Microsoft, and its application scenario is to provide a general solution for quantum programming languages and back-end hardware. SQIR, on the other hand, is an independently developed quantum intermediate representation, which is mainly applied to the verification process of VOQC. The second is the difference in purpose. In our work, the main goal of the formalization of QIR is to provide guarantees for the correctness of quantum programs and the behavior of QIR, while SQIR focuses on the assurance of correctness for quantum circuit optimization. 

In terms of formalizations of quantum programming languages, Singhal {\it et al.}~\cite{qsharpalgo} present $\lambda_{Q\#}$, an idealized version of Q\#. Based on Staton's work~\cite{staton}, they provide a syntactic and semantic formalization of $\lambda_{Q\#}$ that enables it to ensure quantum no-cloning theorem and to provide stacked-like management of qubits. By converting from Q\# to $\lambda_{Q\#}$, additional safety guarantees can be provided for Q\# code. For the Quipper~\cite{green2013quipper} programming language, Mahmoud and Felty~\cite{quipperformal} present Proto-Quipper, which contains the core functionality of Quipper and extends Quipper's system based on the linear specification logic (SL). They also implemented a formalization of Proto-Quipper via Hybird~\cite{hibird}, encoded the complete Proto-Quipper specification, and proved the correctness of its type system. While these works can serve as valuable references for our research, it is critical to recognize significant differences between our work and theirs; our research centers on quantum intermediate representations rather than quantum programming languages.

\section{Concluding Remarks}
\label{sec:conclusion}

In this paper, we have formalized the core functionality of QIR, particularly its abstract syntax and the semantics of its several runtime functions. Our formalization ensures that QIR programs follow the quantum non-cloning theorem and avoid calls to released qubits and arrays. Based on the analysis of real QIR code, we also demonstrated the effectiveness of our formalization in capturing unsafe code in programs.


Our current work is only an initial attempt to formalize the specification of QIR, and more work is needed to refine it. The QIR functions that have been formalized so far are only a small part. More functions need to be formalized, such as $Tuple$ and $String$ in the data type, the semantics of the complete measurement process, and performing batched gate operations on qubits array. From a practical perspective, to apply our formal method directly to the QIR programs for automatic verification (rather than through human analysis as in this work), we need to implement our formal approach with formal interactive theorem checkers such as Coq~\cite{Coq:manual} in further work.

\nocite{*}
\bibliographystyle{eptcs}
\bibliography{generic}

\begin{thebibliography}{10}
\providecommand{\bibitemdeclare}[2]{}
\providecommand{\surnamestart}{}
\providecommand{\surnameend}{}
\providecommand{\urlprefix}{Available at }
\providecommand{\url}[1]{\texttt{#1}}
\providecommand{\href}[2]{\texttt{#2}}
\providecommand{\urlalt}[2]{\href{#1}{#2}}
\providecommand{\doi}[1]{doi:\urlalt{https://doi.org/#1}{#1}}
\providecommand{\eprint}[1]{arXiv:\urlalt{https://arxiv.org/abs/#1}{#1}}
\providecommand{\bibinfo}[2]{#2}

\bibitemdeclare{article}{aleksandrowicz2019qiskit}
\bibitem{aleksandrowicz2019qiskit}
\bibinfo{author}{Gadi \surnamestart Aleksandrowicz\surnameend},
  \bibinfo{author}{Thomas \surnamestart Alexander\surnameend},
  \bibinfo{author}{Panagiotis \surnamestart Barkoutsos\surnameend},
  \bibinfo{author}{Luciano \surnamestart Bello\surnameend},
  \bibinfo{author}{Yael \surnamestart Ben-Haim\surnameend},
  \bibinfo{author}{David \surnamestart Bucher\surnameend},
  \bibinfo{author}{F~Jose \surnamestart Cabrera-Hern{\'a}ndez\surnameend},
  \bibinfo{author}{Jorge \surnamestart Carballo-Franquis\surnameend},
  \bibinfo{author}{Adrian \surnamestart Chen\surnameend},
  \bibinfo{author}{Chun-Fu \surnamestart Chen\surnameend} et~al.
  (\bibinfo{year}{2019}): \emph{\bibinfo{title}{Qiskit: An open-source
  framework for quantum computing}}.
\newblock {\slshape \bibinfo{journal}{Accessed on: Mar}} \bibinfo{volume}{16}.

\bibitemdeclare{misc}{QIR-Alliance}
\bibitem{QIR-Alliance}
\bibinfo{author}{QIR \surnamestart Alliance\surnameend} (\bibinfo{year}{2022}):
  \emph{\bibinfo{title}{{QIR} {A}lliance}}.
\newblock \urlprefix\url{https://www.qir-alliance.org/}.

\bibitemdeclare{inproceedings}{bichsel2020sliq}
\bibitem{bichsel2020sliq}
\bibinfo{author}{Benjamin \surnamestart Bichsel\surnameend},
  \bibinfo{author}{Maximilian \surnamestart Baader\surnameend},
  \bibinfo{author}{Timon \surnamestart Gehr\surnameend} \&
  \bibinfo{author}{Martin \surnamestart Vechev\surnameend}:
  \emph{\bibinfo{title}{Silq: a high-Level quantum language with safe
  uncomputation and intuitive semantics}}.
\newblock In: {\slshape \bibinfo{booktitle}{Proceedings of the 41th ACM SIGPLAN
  Conference on Programming Language Design and Implementation}}.

\bibitemdeclare{manual}{Coq:manual}
\bibitem{Coq:manual}
\bibinfo{author}{The \surnamestart {Coq} {Development}~{Team}\surnameend}
  (\bibinfo{year}{2017}): \emph{\bibinfo{title}{The {Coq} Proof Assistant
  Reference Manual, version 8.7}}.
\newblock \urlprefix\url{http://coq.inria.fr}.

\bibitemdeclare{misc}{OpenQASM}
\bibitem{OpenQASM}
\bibinfo{author}{Andrew~W. \surnamestart Cross\surnameend},
  \bibinfo{author}{Lev~S. \surnamestart Bishop\surnameend},
  \bibinfo{author}{John~A. \surnamestart Smolin\surnameend} \&
  \bibinfo{author}{Jay~M. \surnamestart Gambetta\surnameend}
  (\bibinfo{year}{2017}): \emph{\bibinfo{title}{Open Quantum Assembly
  Language}}, \doi{10.48550/ARXIV.1707.03429}.
\newblock \urlprefix\url{https://arxiv.org/abs/1707.03429}.

\bibitemdeclare{article}{hibird}
\bibitem{hibird}
\bibinfo{author}{Amy~P. \surnamestart Felty\surnameend} \&
  \bibinfo{author}{Alberto \surnamestart Momigliano\surnameend}
  (\bibinfo{year}{2008}): \emph{\bibinfo{title}{Hybrid: {A} Definitional
  Two-Level Approach to Reasoning with Higher-Order Abstract Syntax}}.
\newblock {\slshape \bibinfo{journal}{CoRR}} \bibinfo{volume}{abs/0811.4367}.
\newblock \eprint{0811.4367}.

\bibitemdeclare{misc}{QIR}
\bibitem{QIR}
\bibinfo{author}{Alan \surnamestart Geller\surnameend} (\bibinfo{year}{2020}):
  \emph{\bibinfo{title}{Introducing Quantum Intermediate Representation
  {(QIR)}}}.
\newblock
  \urlprefix\url{https://devblogs.microsoft.com/qsharp/introducing-quantum-intermediate-representation-qir/}.

\bibitemdeclare{inproceedings}{green2013quipper}
\bibitem{green2013quipper}
\bibinfo{author}{Alexander~S \surnamestart Green\surnameend},
  \bibinfo{author}{Peter~LeFanu \surnamestart Lumsdaine\surnameend},
  \bibinfo{author}{Neil~J \surnamestart Ross\surnameend},
  \bibinfo{author}{Peter \surnamestart Selinger\surnameend} \&
  \bibinfo{author}{Beno{\^\i}t \surnamestart Valiron\surnameend}
  (\bibinfo{year}{2013}): \emph{\bibinfo{title}{Quipper: a scalable quantum
  programming language}}.
\newblock In: {\slshape \bibinfo{booktitle}{Proceedings of the 34th ACM SIGPLAN
  conference on Programming language design and implementation}}, pp.
  \bibinfo{pages}{333--342}.

\bibitemdeclare{article}{SQIR}
\bibitem{SQIR}
\bibinfo{author}{Kesha \surnamestart Hietala\surnameend},
  \bibinfo{author}{Robert \surnamestart Rand\surnameend},
  \bibinfo{author}{Shih-Han \surnamestart Hung\surnameend},
  \bibinfo{author}{Xiaodi \surnamestart Wu\surnameend} \&
  \bibinfo{author}{Michael \surnamestart Hicks\surnameend}
  (\bibinfo{year}{2021}): \emph{\bibinfo{title}{A Verified Optimizer for
  Quantum Circuits}}.
\newblock {\slshape \bibinfo{journal}{Proc. ACM Program. Lang.}}
  \bibinfo{volume}{5}(\bibinfo{number}{POPL}), \doi{10.1145/3434318}.
\newblock \urlprefix\url{https://doi.org/10.1145/3434318}.

\bibitemdeclare{inproceedings}{LLVM:CGO04}
\bibitem{LLVM:CGO04}
\bibinfo{author}{Chris \surnamestart Lattner\surnameend} \&
  \bibinfo{author}{Vikram \surnamestart Adve\surnameend}
  (\bibinfo{year}{2004}): \emph{\bibinfo{title}{{LLVM}: A Compilation Framework
  for Lifelong Program Analysis and Transformation}}.
\newblock \bibinfo{address}{San Jose, CA, USA}, pp. \bibinfo{pages}{75--88}.

\bibitemdeclare{inproceedings}{mlir}
\bibitem{mlir}
\bibinfo{author}{Chris \surnamestart Lattner\surnameend},
  \bibinfo{author}{Mehdi \surnamestart Amini\surnameend}, \bibinfo{author}{Uday
  \surnamestart Bondhugula\surnameend}, \bibinfo{author}{Albert \surnamestart
  Cohen\surnameend}, \bibinfo{author}{Andy \surnamestart Davis\surnameend},
  \bibinfo{author}{Jacques \surnamestart Pienaar\surnameend},
  \bibinfo{author}{River \surnamestart Riddle\surnameend},
  \bibinfo{author}{Tatiana \surnamestart Shpeisman\surnameend},
  \bibinfo{author}{Nicolas \surnamestart Vasilache\surnameend} \&
  \bibinfo{author}{Oleksandr \surnamestart Zinenko\surnameend}
  (\bibinfo{year}{2021}): \emph{\bibinfo{title}{{{MLIR}}: Scaling Compiler
  Infrastructure for Domain Specific Computation}}.
\newblock In: {\slshape \bibinfo{booktitle}{2021 {{IEEE/ACM}} International
  Symposium on Code Generation and Optimization (CGO)}}, pp.
  \bibinfo{pages}{2--14}, \doi{10.1109/CGO51591.2021.9370308}.

\bibitemdeclare{inproceedings}{k-llvm}
\bibitem{k-llvm}
\bibinfo{author}{Liyi \surnamestart Li\surnameend} \& \bibinfo{author}{Elsa~L.
  \surnamestart Gunter\surnameend} (\bibinfo{year}{2020}):
  \emph{\bibinfo{title}{{K-LLVM: A Relatively Complete Semantics of LLVM IR}}}.
\newblock In \bibinfo{editor}{Robert \surnamestart Hirschfeld\surnameend} \&
  \bibinfo{editor}{Tobias \surnamestart Pape\surnameend}, editors: {\slshape
  \bibinfo{booktitle}{34th European Conference on Object-Oriented Programming
  (ECOOP 2020)}}, {\slshape \bibinfo{series}{Leibniz International Proceedings
  in Informatics (LIPIcs)}} \bibinfo{volume}{166}, \bibinfo{publisher}{Schloss
  Dagstuhl--Leibniz-Zentrum f{\"u}r Informatik}, \bibinfo{address}{Dagstuhl,
  Germany}, pp. \bibinfo{pages}{7:1--7:29}, \doi{10.4230/LIPIcs.ECOOP.2020.7}.
\newblock \urlprefix\url{https://drops.dagstuhl.de/opus/volltexte/2020/13164}.

\bibitemdeclare{article}{quipperformal}
\bibitem{quipperformal}
\bibinfo{author}{Mohamed~Yousri \surnamestart Mahmoud\surnameend} \&
  \bibinfo{author}{Amy~P. \surnamestart Felty\surnameend}
  (\bibinfo{year}{2018}): \emph{\bibinfo{title}{Formalization of Metatheory of
  the Quipper Quantum Programming Language in a Linear Logic}}.
\newblock {\slshape \bibinfo{journal}{CoRR}} \bibinfo{volume}{abs/1812.03624}.
\newblock \eprint{1812.03624}.

\bibitemdeclare{misc}{MLIR-Quantum}
\bibitem{MLIR-Quantum}
\bibinfo{author}{Alexander \surnamestart McCaskey\surnameend} \&
  \bibinfo{author}{Thien \surnamestart Nguyen\surnameend}
  (\bibinfo{year}{2021}): \emph{\bibinfo{title}{A {MLIR} Dialect for Quantum
  Assembly Languages}}, \doi{10.48550/ARXIV.2101.11365}.
\newblock \urlprefix\url{https://arxiv.org/abs/2101.11365}.

\bibitemdeclare{}{qir-spec}
\bibitem{qir-spec}
\bibinfo{author}{\surnamestart Microsoft\surnameend} (\bibinfo{year}{2022}):
  \emph{\bibinfo{title}{Quantum Intermediate Representation (QIR)}}.
\newblock
  \urlprefix\url{https://github.com/qir-alliance/qir-spec/tree/main/specification}.

\bibitemdeclare{article}{K}
\bibitem{K}
\bibinfo{author}{D.~\surnamestart Monderer\surnameend} \&
  \bibinfo{author}{M.~\surnamestart Tennenholtz\surnameend}
  (\bibinfo{year}{2004}): \emph{\bibinfo{title}{K-Implementation}}.
\newblock {\slshape \bibinfo{journal}{Journal of Artificial Intelligence
  Research}} \bibinfo{volume}{21}, pp. \bibinfo{pages}{37--62},
  \doi{10.1613/jair.1231}.
\newblock \urlprefix\url{https://doi.org/10.1613\%2Fjair.1231}.

\bibitemdeclare{inproceedings}{softbound}
\bibitem{softbound}
\bibinfo{author}{Santosh \surnamestart Nagarakatte\surnameend},
  \bibinfo{author}{Jianzhou \surnamestart Zhao\surnameend},
  \bibinfo{author}{Milo~M.K. \surnamestart Martin\surnameend} \&
  \bibinfo{author}{Steve \surnamestart Zdancewic\surnameend}
  (\bibinfo{year}{2009}): \emph{\bibinfo{title}{SoftBound: Highly Compatible
  and Complete Spatial Memory Safety for c}}.
\newblock In: {\slshape \bibinfo{booktitle}{Proceedings of the 30th ACM SIGPLAN
  Conference on Programming Language Design and Implementation}},
  \bibinfo{series}{PLDI '09}, \bibinfo{publisher}{Association for Computing
  Machinery}, \bibinfo{address}{New York, NY, USA}, p.
  \bibinfo{pages}{245–258}, \doi{10.1145/1542476.1542504}.
\newblock \urlprefix\url{https://doi.org/10.1145/1542476.1542504}.

\bibitemdeclare{article}{10.1145/1543135.1542504}
\bibitem{10.1145/1543135.1542504}
\bibinfo{author}{Santosh \surnamestart Nagarakatte\surnameend},
  \bibinfo{author}{Jianzhou \surnamestart Zhao\surnameend},
  \bibinfo{author}{Milo~M.K. \surnamestart Martin\surnameend} \&
  \bibinfo{author}{Steve \surnamestart Zdancewic\surnameend}
  (\bibinfo{year}{2009}): \emph{\bibinfo{title}{SoftBound: Highly Compatible
  and Complete Spatial Memory Safety for c}}.
\newblock {\slshape \bibinfo{journal}{SIGPLAN Not.}}
  \bibinfo{volume}{44}(\bibinfo{number}{6}), p. \bibinfo{pages}{245–258},
  \doi{10.1145/1543135.1542504}.
\newblock \urlprefix\url{https://doi.org/10.1145/1543135.1542504}.

\bibitemdeclare{article}{qcor}
\bibitem{qcor}
\bibinfo{author}{Thien \surnamestart {Nguyen}\surnameend},
  \bibinfo{author}{Anthony \surnamestart {Santana}\surnameend},
  \bibinfo{author}{Tyler \surnamestart {Kharazi}\surnameend},
  \bibinfo{author}{Daniel \surnamestart {Claudino}\surnameend},
  \bibinfo{author}{Hal \surnamestart {Finkel}\surnameend} \&
  \bibinfo{author}{Alexander \surnamestart {McCaskey}\surnameend}
  (\bibinfo{year}{2020}): \emph{\bibinfo{title}{{Extending C++ for
  Heterogeneous Quantum-Classical Computing}}}.
\newblock {\slshape \bibinfo{journal}{arXiv
  e-prints}}:\bibinfo{eid}{arXiv:2010.03935}.
\newblock \eprint{2010.03935}.

\bibitemdeclare{misc}{nielsen2002quantum}
\bibitem{nielsen2002quantum}
\bibinfo{author}{Michael~A \surnamestart Nielsen\surnameend} \&
  \bibinfo{author}{Isaac \surnamestart Chuang\surnameend}
  (\bibinfo{year}{2002}): \emph{\bibinfo{title}{Quantum computation and quantum
  information}}.

\bibitemdeclare{article}{preskill2018quantum}
\bibitem{preskill2018quantum}
\bibinfo{author}{John \surnamestart Preskill\surnameend}
  (\bibinfo{year}{2018}): \emph{\bibinfo{title}{Quantum computing in the {NISQ}
  era and beyond}}.
\newblock {\slshape \bibinfo{journal}{Quantum}} \bibinfo{volume}{2},
  p.~\bibinfo{pages}{79}.

\bibitemdeclare{article}{ibm2021qiskit}
\bibitem{ibm2021qiskit}
\bibinfo{author}{IBM \surnamestart Research\surnameend}:
  \emph{\bibinfo{title}{Qiskit}}.
\newblock {\slshape \bibinfo{journal}{Accessed on: June, 2021}}.
\newblock \urlprefix\url{https://qiskit.org}.

\bibitemdeclare{misc}{qsharpalgo}
\bibitem{qsharpalgo}
\bibinfo{author}{Kartik \surnamestart Singhal\surnameend},
  \bibinfo{author}{Kesha \surnamestart Hietala\surnameend},
  \bibinfo{author}{Sarah \surnamestart Marshall\surnameend} \&
  \bibinfo{author}{Robert \surnamestart Rand\surnameend}
  (\bibinfo{year}{2022}): \emph{\bibinfo{title}{Q\# as a Quantum Algorithmic
  Language}}, \doi{10.48550/ARXIV.2206.03532}.
\newblock \urlprefix\url{https://arxiv.org/abs/2206.03532}.

\bibitemdeclare{article}{staton}
\bibitem{staton}
\bibinfo{author}{Sam \surnamestart Staton\surnameend} (\bibinfo{year}{2015}):
  \emph{\bibinfo{title}{Algebraic Effects, Linearity, and Quantum Programming
  Languages}}.
\newblock {\slshape \bibinfo{journal}{SIGPLAN Not.}}
  \bibinfo{volume}{50}(\bibinfo{number}{1}), p. \bibinfo{pages}{395–406},
  \doi{10.1145/2775051.2676999}.
\newblock \urlprefix\url{https://doi.org/10.1145/2775051.2676999}.

\bibitemdeclare{inproceedings}{10.1145/2676726.2676999}
\bibitem{10.1145/2676726.2676999}
\bibinfo{author}{Sam \surnamestart Staton\surnameend} (\bibinfo{year}{2015}):
  \emph{\bibinfo{title}{Algebraic Effects, Linearity, and Quantum Programming
  Languages}}.
\newblock In: {\slshape \bibinfo{booktitle}{Proceedings of the 42nd Annual ACM
  SIGPLAN-SIGACT Symposium on Principles of Programming Languages}},
  \bibinfo{series}{POPL '15}, \bibinfo{publisher}{Association for Computing
  Machinery}, \bibinfo{address}{New York, NY, USA}, p.
  \bibinfo{pages}{395–406}, \doi{10.1145/2676726.2676999}.
\newblock \urlprefix\url{https://doi.org/10.1145/2676726.2676999}.

\bibitemdeclare{inproceedings}{svore2018q}
\bibitem{svore2018q}
\bibinfo{author}{Krysta \surnamestart Svore\surnameend}, \bibinfo{author}{Alan
  \surnamestart Geller\surnameend}, \bibinfo{author}{Matthias \surnamestart
  Troyer\surnameend}, \bibinfo{author}{John \surnamestart Azariah\surnameend},
  \bibinfo{author}{Christopher \surnamestart Granade\surnameend},
  \bibinfo{author}{Bettina \surnamestart Heim\surnameend},
  \bibinfo{author}{Vadym \surnamestart Kliuchnikov\surnameend},
  \bibinfo{author}{Mariia \surnamestart Mykhailova\surnameend},
  \bibinfo{author}{Andres \surnamestart Paz\surnameend} \&
  \bibinfo{author}{Martin \surnamestart Roetteler\surnameend}
  (\bibinfo{year}{2018}): \emph{\bibinfo{title}{Q\#: enabling scalable quantum
  computing and development with a high-level DSL}}.
\newblock In: {\slshape \bibinfo{booktitle}{Proceedings of the Real World
  Domain Specific Languages Workshop 2018}}, pp. \bibinfo{pages}{1--10}.

\bibitemdeclare{article}{cirq2018google}
\bibitem{cirq2018google}
\bibinfo{author}{Google AI~Quantum \surnamestart team\surnameend}
  (\bibinfo{year}{2018}): \emph{\bibinfo{title}{Cirq}}.
\newblock \urlprefix\url{https://github.com/quantumlib/Cirq}.

\bibitemdeclare{article}{projectq2017projectq}
\bibitem{projectq2017projectq}
\bibinfo{author}{Project{Q} \surnamestart Team\surnameend}
  (\bibinfo{year}{2017}): \emph{\bibinfo{title}{Project{Q}}}.
\newblock {\slshape \bibinfo{journal}{Accessed on: April, 2020}}.
\newblock \urlprefix\url{https://projectq.ch/}.

\bibitemdeclare{inproceedings}{vellvm}
\bibitem{vellvm}
\bibinfo{author}{Jianzhou \surnamestart Zhao\surnameend},
  \bibinfo{author}{Santosh \surnamestart Nagarakatte\surnameend},
  \bibinfo{author}{Milo~M.K. \surnamestart Martin\surnameend} \&
  \bibinfo{author}{Steve \surnamestart Zdancewic\surnameend}
  (\bibinfo{year}{2012}): \emph{\bibinfo{title}{Formalizing the {LLVM}
  Intermediate Representation for Verified Program Transformations}}.
\newblock In: {\slshape \bibinfo{booktitle}{Proceedings of the 39th Annual ACM
  SIGPLAN-SIGACT Symposium on Principles of Programming Languages}},
  \bibinfo{series}{POPL '12}, \bibinfo{publisher}{Association for Computing
  Machinery}, \bibinfo{address}{New York, NY, USA}, p.
  \bibinfo{pages}{427–440}, \doi{10.1145/2103656.2103709}.
\newblock \urlprefix\url{https://doi.org/10.1145/2103656.2103709}.

\end{thebibliography}
\end{document}